\documentclass[10pt,a4paper]{article}
\usepackage[latin1]{inputenc}
\usepackage[english]{babel}
\usepackage{amsmath}
\usepackage{amsfonts}
\usepackage{amssymb}
\usepackage{makeidx}
\usepackage{graphicx}
\usepackage[normalem]{ulem}
\usepackage[left=2cm,right=2cm,top=2cm,bottom=2cm]{geometry}
\usepackage{color}

\usepackage{amssymb,amsthm,xspace}
\usepackage{latexsym} 
\usepackage{lineno}

\theoremstyle{remark}

\newtheorem{proposition}{Proposition}

\theoremstyle{definition}

\author{Alberto Pasanisi*, C\^{o}me Roero**\\ Nicolas Bousquet***\\ Emmanuel Remy***\\
** EDF R\&D, European Institute for Energy Research\\ Emmy-Noether-Str. 11, 76131, Kalrsruhe, Germany \\
** INRIA - University Paris-Sud, Mathematics Dept.\\ Bat. 425, 91405 Orsay, France\\
*** EDF R\&D, Industrial Risk Management Dept. \\ 6 quai Watier, 78401 Chatou, France}
\title{On the practical interest of discrete Inverse P\'{o}lya and Weibull-1 models in industrial reliability studies}

\begin{document}
\maketitle
\begin{abstract}
Engineers often cope with the problem of assessing the lifetime of industrial components, under the basis of observed industrial feedback data. Usually, lifetime is modelled as a continuous random variable, for instance exponentially or Weibull distributed. However, in some cases, the features of the piece of equipment under investigation rather suggest the use of discrete probabilistic models. This happens for an equipment which only operates on cycles or on demand. In these cases, the lifetime is rather measured in number of cycles or number of demands before failure, therefore, in theory, discrete models should be more appropriate. This article aims at bringing some light to the practical interest for the reliability engineer in using two discrete models among the most popular: the Inverse P\'{o}lya  distribution (IPD), based on a P\'{o}lya urn scheme, and the so-called Weibull-1 (W1) model. It is showed that, for different reasons, the practical use of both models should be restricted to specific industrial situations. 
In particular, when nothing is a priori known over the nature of ageing and/or data are heavily right-censored, they can remain of limited interest with respect to more flexible continuous lifetime models such as the usual Weibull distribution. Nonetheless, the intuitive meaning given to the IPD distribution favors its use by engineers in low (decelerated) ageing situations.\\


\end{abstract}

\section{Introduction}\label{sec:intro}
The use of discrete survival models is naturally considered when the lifetime of the piece of equipment under investigation cannot be properly expressed as a calendar time. It is the case of components which do not operate continuously and are only occasionally solicited. On-off switch or auxiliary power devices are typical examples: the activations of the switch or the starts of the engine can be considered as ''occasional stresses" or solicitations for the equipment. In these cases, for reliability assessment purposes, the variable characterizing the lifetime of the component is not the operating time (e.g. measured in hours), but rather the number $N$ of solicitations that the component can bear before failure. Another case of discrete lifetime data concerns pieces of equipment which only operate on cycles and the collected information is just the correct (or incorrect) behaviour at a given cycle.
In both cases, the problem can be formalized by stating that the variable "lifetime $N$", considered as random, is discrete.


In spite of the potential amount of case-studies in which discrete lifetime models could be considered as appealing tools for the engineers, they have been relatively not much investigated (theoretically and/or practically) in comparison with continuous models, which are nowadays largely used in industrial practice. 
The first scientific article proposing a discrete lifetime model (derived from the continuous Weibull distribution) dates from the mid of the 70's \cite{Nakagawa1975}. Bracquemond and Gaudoin, in their seminal work, \cite{Bracquemond2003}, provide a quite exhaustive review of discrete distributions for lifetime data, including numerous references. Roughly speaking, they can be grouped into two categories: the ones derived from continuous models and those derived from urn schemes.  After a concise statistical study, for various practical reasons and because of their convenient properties, the authors recommend in most situations the use of the Type I discrete Weibull family, also called Weibull-1 (W1), first defined by Nakagawa and Osaki \cite{Nakagawa1975}, or the Eggenberger-P\'{o}lya distribution \cite{Eggenberger1923}.  In a more engineering framework, Clarotti et al. \cite{Clarotti1997} stressed the importance  to dispose of intuitive models, in the sense that their features have appealing meanings for the (often non-statistician) practitioner and can be intuitively interpreted by experts. Consequently, they considered that the Inverse P\'{o}lya distribution (IPD) can be especially valuable in ageing problems. This offers an alternative solution to the difficulty of deriving discrete distributions from continuous ones, highlighted, for instance, by Lai \cite{Lai2013}.

When conducting risk and reliability analyses and eliciting one or several lifetime models, engineers ordinarily try to reach a trade-off between intuition (or practical interpretation), versatility and flexibility in use. 
Especially in situations where such analyses must be submitted to control authorities, e.g. when they are conducted by polluting industries about their production components, the relevance of these models must be demonstrated. To avoid possible cumbersome consequences in terms of safety and availability, several features are to be checked, that connect qualitative and quantitative aspects of the behavior of the component. The versatility of the model can indeed be interpreted as its ability to represent several possible states of a life cycle  in running conditions, as rejuvenation (due to preventive maintenance) or degradation due to ageing (which, as a function of time, can be accelerated, stable, or decelerated). Such properties are intrinsically connected to the shape of the hazard rate function. For discrete models, the hazard rate, denoted $\lambda(n)$ in the remainder of this article, is classically defined as the conditional probability of failure after a given number $n$ of solicitations, given that the component has already beared  $n-1$ solicitations. Derived from the definition of the hazard function in the continuous case (where $\lambda$ can be interpreted as a "conditional density function"~\cite{Lai2013}), it is actually the most popular one in the reliability community (e.g. \cite{Salvia1982, Shaked1995, Lawless2003, Rinne2008, Almalki2013}). For instance, a concave hazard rate, characterized by a negative second-order derivative of $\lambda(n)$, describe situations of decelerated ageing. Such situations can be reflected in the failure data collection since good maintenance strategies can limit the effect of ageing. However, one could choose to take into account the possibility of a rapid degradation of the component in given running conditions, and therefore select a class of more flexible models. Note that an alternative definition of the hazard rate, defended in \cite{Roy1992} and discussed by~\cite{Xie2002, Lai2013}, is based on the logarithm of the ratio of the survival function in $n-1$ and $n$, but it will be not used in this article. 

Furthermore, the selected models must be accompanied with some clear and practical rule of inference from feedback experience data (FED), obtained from tests or in situ measurements. Such data can rarely be considered as identically and independently distributed because of the experimental limitations. These observed limitations are often traduced statistically as censoring experiments, which constitute sometimes the major part of the datasets. A huge rate of right-censoring typically affects  the FED related to highly reliable components (ie.,  that fail after a relatively high number of solicitations) of power producers and suppliers.

Among them, \'Electricit\'e de France (EDF) is obviously concerned with the lifetime assessment of the components of its power plants as well as its transmission and distribution facilities, and encounters frequently the problem of selecting and assessing relevant models for equipments that can fail on demand. Testing the relevance of the IPD and W1 models to such contexts, reflected by censored FED samples, is of practical interest for many other industries (e.g, in fields concerned with important safety constraints) and motivates the present work. This article therefore aims at providing guidelines and advices for the use of both models in concrete situations, at the light of several experiments conducted on simulated and real datasets.

The remainder of this article is organized as follows. The origin, definition and properties of examined models are presented in Sections \ref{sec:Polya} and \ref{sec:Weibull1Intro}. It is worth noting that in Appendix details are given on a maximum likelihood estimation method for the IPD model, which was lacking in the reliability community to popularize the use of this model (while the estimation of the W1 model is well known, cf. \cite{Bracquemond2003}).  Each section incorporates a discussion of the own merits and limitations of the considered model. Especially, in spite of its appealing simplicity and clarity of interpretation by the engineers' viewpoint, it is pointed out that the IPD model is only adapted to situations where ageing is low and decelerated. On the other hand, the properties of the W1 model may sound odd for the practitioners and make the interpretation of its main features quite complicated.  
In Section  \ref{sec:Weibullcont}  the closeness between the W1 and the usual (continuous) Weibull distributions is formally proven, which pleads for using the latter as a robust and convenient approximation of the discrete model in the engineering practice. 
A part of this section is dedicated to the numerical investigation of the properties and limitations of each considered model, using simulated and real datasets. Its results support the choice of the continuous Weibull distribution as a practical proxy of the W1 distribution. Finally, in Section \ref{sec:real} the selected models are fitted over two real industrial datasets  of component lifetime, and a study of the nature of ageing is conducted. 
A discussion section ends this article by sketching the main teachings of this study - how and when the two models can be relevant - and proposing some avenues for further research.

\section{The Inverse P\'{o}lya (IPD) model}\label{sec:Polya}

\subsection{Origin, definition and main features}

The use of urn sampling schemes to model the behaviour of living \cite{Marshall1993, Ivanova2000} or industrial~\cite{Alajaji1993, Bracquemond2001} systems has often been considered.
The basic principle of the numerous probabilistic models based on the P\'{o}lya urn scheme~\cite{Bracquemond2003,Johnson2005,Mahmoud2008}, first introduced in the 20's of the last century \cite{Eggenberger1923,Polya1930}, is to consider an urn in which, at the beginning of the experience, there are $a$ black balls and $b$ red ones, so that the probability to extract a black ball after a random trial is $a/(a+b)$. If a red ball is sampled, then $z$ new black balls are added (together with the sampled red ball), thus increasing the probability to sample a black ball.
This scheme suggests an appealing probabilistic lifetime model \cite{Clarotti1997} for discrete data: each solicitation of the piece of equipment is considered as a trial in a P\'{o}lya urn, where black balls are associated to the event "failure" and red ones to the event "correctly operating". Adding new black balls can easily be interpreted as the result of a deterioration process (summarized by the term ''ageing").

The random variable $N$ "number of the trial at which the failure occurs" follows a so-called Inverse P\'{o}lya distribution (IPD), namely $N \sim \text{IPD} ( \alpha,\zeta)$
where, following the parameterization choices made in \cite{Bracquemond2003},
\begin{equation*}
\alpha=\frac{a}{a+b} ~~\text{and}~~\zeta=\frac{z}{a+b},~~~\text{with } 0<\alpha<1 ~~\text{and}~~ \zeta >0.
\end{equation*}

Notice that $\alpha$  can be easily interpreted as the probability of failure at the first solicitation ($n=1$).  The nature of ageing of the component governs the value of the parameter $\zeta$: the higher $\zeta$, the more severe will be the ageing. The expression of the main reliability quantities of interest are given below. 
\begin{equation}\label{eq:SurvivalAndHazardPolya} 
\begin{cases} 
\text{Hazard rate:~~~} \lambda(n)=\mathbb{P}\left[ N=n|N>n-1 \right] = \dfrac{\alpha+(n-1)\zeta}{1+(n-1)\zeta} \\
\\
\text{Probability of failure after $n$ solicitations:~} p(n)=\mathbb{P}\left[N=n \right] = \dfrac{(1-\alpha)^{n-1} (\alpha+(n-1)\zeta)}{\prod_{i=1}^{i=n}(1+(i-1))\zeta}\\
\\
\text{Survival function:~} S(n)=\mathbb{P}\left[N>n\right]=\dfrac{(1-\alpha)^n}{\prod_{i=1}^{i=n}(i+(i-1))\zeta}\\
\\
\text{MTTF:~} \mathbb{E}\left[N\right]=\dfrac{(1-\zeta)\zeta^{(1/\zeta-2)}}{(1-\beta)^{(1-\zeta)/\zeta}} \exp \left(\dfrac{1-\alpha}{\zeta}\right)\gamma\left( \dfrac{1-\zeta}{\zeta} , \dfrac{1-\alpha}{\zeta} \right)
\end{cases}
\end{equation}

In the expression of the Mean Time To Failure (MTTF) above, $\gamma(\cdot , \cdot )$ is the so-called lower incomplete Gamma function:
\begin{equation*}
\gamma(u,v)=\int_{0}^{v} x^{u-1} \exp(-x) \, dx .
\end{equation*}
As an increasing function of $n$, the failure rate $\lambda(n)$ makes the model likely to represent some ageing behaviors. 


\subsection{Modelling ageing using IPD: a major limitation}\label{sec:PolyaAgeing}

The ageing of the component under investigation can be assumed by ensuring the condition $\zeta>0$. However, in practical applications one is also interested in describing situations where the ageing increases or decreases as the observed lifetime $N$ 
increases. This issue is solved by studying the sign of the second-order derivative of the failure rate. In the case of the P\'{o}lya model the second-order discrete derivative of $\lambda(n)$  can be written, for $n>2$ (after some algebra):
\begin{eqnarray}
\lambda''(n) & = & 
\lambda(n)-2\lambda(n-1)+\lambda(n-2) \nonumber  \\ 
& = & \frac{2(\alpha-1)\zeta^{2}}{(1+(n-1)\zeta)(1+(n-2)\zeta)(1+(n-3)\zeta)}. \label{eq:FailureRatePolya2ndDeriv}
\end{eqnarray}
Under the conditions: $\alpha<1$, $\zeta>0$ and $n>2$, it is easy to verify that the numerator and the denominator of Equation~\ref{eq:FailureRatePolya2ndDeriv} are negative and positive, respectively. Thus, for any value of $\zeta$, the second-order derivative of the failure rate is negative, namely the IPD can only model situations of decelerated ageing.  This result  is confirmed by the intuition: if at each solicitation a number $z$ of black balls is added into the urn, $n$ increases progressively and the number $z$ of added balls becomes smaller  than the number of the black balls already present. 
For large values of $n$, $z$ becomes negligible and the added balls do not influence the failure probability anymore. This situation can occur when a preventive maintenance is sufficient enough to prevent close breakdowns, that typically herald the close end of a component lifetime. For this reason, and because the meanings of its parameters are rather intuitive, the Inverse P\'olya model deserves interest in the reliability and risk community, although its use must also be strictly reserved to low ageing components or systems. \\

Due to the non-trivial handling of IPD, numerical computations proposed in the remainder needed to dispose of methods for simulating datasets and estimating the parameters in realistic cases. More precisely, it is needed: 
\begin{itemize}
\item to have a view of the range of realistic values for $(\alpha,\zeta)$, associated to various ageing situations; 
\item to describe a sampling method, given $(\alpha,\zeta)$: this is done in Appendix;
\item to describe an appropriate estimation method; a maximum likelihood (ML) method devoted to this task is presented in Appendix too.
\end{itemize}
An answer to the first item is provided by the experiment resumed in Table~\ref{tab:test-weibull-polya}. It is inspired by the case of both continuous and discrete Weibull distributions, in which the shape parameter $\beta$ appears as an immediate indicator of the nature of ageing (see also Section~\ref{sec:Weibull1Ageing}); in this case, its value can help the reliability engineer to synthesize the behavior of the considered component. Analogously, it is therefore wanted to simply characterize the nature of ageing for the inverse P\'{o}lya model. In a non-exhaustive way, several situations can be simulated using Weibull samples, on which inverse P\'{o}lya models are then fitted. On Table \ref{tab:test-weibull-polya}, a range of such situations, from rejuvenation to accelerated ageing, are considered.
In practice, the values of $\zeta/\alpha$ shown in this table have been obtained by fitting IPD on a number of (discretized) lifetimes sampled from the usual (continuous) Weibull distribution.

Apart from providing ranges of plausible values for the ratio $\zeta/\alpha$ in presence of rejuvenation or soft ageing, these results highlight the fact that, following engineering common sense, finding an estimate of this ratio upper than one discredits the "physical" relevance of the inverse P\'{o}lya model. Actually, a model considering that at each solicitation, the reliability decreases of an amount greater than the initial reliability, although mathematically possible, seems not coherent  from an engineering viewpoint.

\begin{table}[hbtp]
\centering
\begin{tabular}{llll}
 Scenario & Weibull shape parameter $\beta$ && Ratio $\zeta/\alpha$ \\
 \hline
rejuvenation & $\beta\leq 0.9$ && $\leq 10^{-5}$ \\
no rejuvenation / no ageing  & $\beta=1$ && $[8.10^{-5},10^{-4}]$\\
soft decelerated ageing & $\beta=1.2$ && $[5.8.10^{-4},7.10^{-4}]$ \\
classical decelerated ageing (1) & $\beta=1.5$ &&  $[2.6.10^{-3},3.2.10^{-3}]$\\
classical decelerated ageing (2) & $\beta=1.8$ && $[2.10^{-2},4.10^{-2}]$ \\
non-accelerated ageing & $\beta=2$ && $[0.25,0.35]$\\
accelerated ageing  & $\beta=2.25$ && $[1.28,1.35]$\\
strongly accelerated ageing  & $\beta=2.5$ && $[1.48,1.85]$ \\
 \hline
\end{tabular}
\caption{\label{tab:test-weibull-polya} \small Typical magnitudes for the ratio $\zeta/\alpha$ as a function of a Weibull shape parameter $\beta$, that indicates qualitatively the ageing behaviour of a component. These ranges of values were estimated by ML estimation from 500 discretized Weibull samples of size 1000, generated using scale parameter values in $\{10,100,500,1000\}$.} 
\end{table}

Except in situations of low aging, where the FED is composed with data reflecting both the degradation and maintenance process affecting the component or system under investigation, the irrelevance of the IPD model can be a serious concern for reliability engineers in several cases. For instance, when the degradation of similar components can be fast and occur in a small interval of time, despite possible maintenance efforts. Possible causes of accelerated ageing can be brutal changes in the environnement. 
Another typical case is when the considered maintenance is only corrective (namely, when a single replacement is made after a failure). 
The poor predictive properties of the IPD model in presence of data concerning systems presenting an increasing failure rate are exemplified later in the text (cf. Figure \ref{fig:SimulatedHazardAndData}). For this reason, the remainder of this chapter is mostly focused on another popular probabilistic model for discrete lifetimes, derived from the continuous Weibull distribution.



\section{The Weibull-1 (W1) model}\label{sec:Weibull1Intro}

\subsection{Origin, definition and main features}

The Weibull distribution 
 is the most popular probabilistic model for continuous lifetime data in engineering. Several discrete versions of the  Weibull model for discrete data have been proposed.
The so-called "Weibull-1" $\text{W}_1 ( \eta,\beta)$ distribution (or Type I Weibull distribution), which is historically the first one, was proposed in 1975~\cite{Nakagawa1975} and is recommended by several authors \cite{Bracquemond2003}. It can be derived from the usual (continuous) Weibull distribution by time discretization \cite{Lai2013ASMBI} or alternatively defined by means of its survival function, which has formally the same expression as the continuous Weibull's one. Thus, the following notations and definitions apply:

\begin{equation}\label{eq:SurvivalAndHazardWeibull1}
\begin{cases}
\text{Hazard function:~~~} \lambda(n)=1- \exp \left[ - \left( \dfrac{n}{\eta}\right)^\beta + \left( \dfrac{n-1}{\eta} \right)^\beta   \right] \\
\\
\text{Prob. of failure after $n$ solicitations:~} p(n) = \exp \left[ - \left( \dfrac{n-1}{\eta}\right)^\beta \right] -  \exp \left[ - \left( \dfrac{n}{\eta} \right) ^\beta \right] \\
\\
\text{Survival function:~~~} S(n)=\exp\left[- \left( \dfrac{n}{\eta} \right)^\beta     \right].
\end{cases}
\end{equation}
Although no closed form of the MTTF exists for the W1 model, upper and lower bounds are provided in Proposition \ref{prop1} (page~\pageref{prop1}).
The W1 model is often re-parametrized as  $\text{W}_1(\theta ,\beta)$, with $\theta=\exp \left( -1/ \eta ^\beta \right)$. This parametrization allows a very easy interpretation of the parameter $\theta$: actually, $1-\theta$ is the probability of failure at the first solicitation (i.e. for $n=1$). Nevertheless, the advantage of the parametrization $(\eta,\beta)$ is the easiness of the comparison between the two distributions $\text{W}_1(\eta,\beta)$ and  $\text{W}(\eta,\beta)$ (i.e. W1 and continuous Weibull having the same parameters) and will therefore be used in the remainder of this article. This comparison will be conducted in Section \ref{sec:Weibullcont}.

Other discrete distributions can be derived from the continuous Weibull one, especially the Weibull-2 distribution \cite{Stein1984}, which preserves the power function form of the hazard rate, and the Weibull-3 distribution~\cite{Padgett1985}. See also~\cite{Jazi2010, Alzaatreh2012, Bebbington2012, Lai2013} for examples of more complex related distributions, as well as the recent review article of \cite{Almalki2013} which proposes several variants of both discrete and continuous Weibull distributions.
However, as reminded in~\cite{Rinne2008}, no discrete distribution exists that can mimic all the functional forms and the properties, so familiar to engineers, of the continuous Weibull one. 

Although some uses of the W1 distribution can be found in the reliability literature, e.g. modelling the number of shocks~\cite{Sheu1998,Sheu1998EJOR} and the number of preventive maintenance actions~\cite{Liao2009, Liao2011} before failure, as well as the number of items produced in an in-control state of a manufacturing process before shifting to an out-of-control state~\cite{Wang2001, Wang2003, Wang2009, Tsai2011},  this model still remains little used in the engineering practice. It is worth noting that it had some success outside the industrial reliability context. It has been used for instance for modelling discretized durations of wind events~\cite{Castino1998}, recruitments of trees in \textit{Growth and yield} models of forests~\cite{Fortin2009}, survival times of individuals affected by infectious diseases \cite{Reich2012}, microbiological counts in drinking water \cite{Englehardt2011}, the distribution of polymeric particles hosting the active agent in drug release experiments~\cite{Grassi2000} or the number of cells population doublings until senescence in \textit{in vitro} experiments~\cite{Wein2001}. 





\subsection{Modelling ageing using the W1 distribution}\label{sec:Weibull1Ageing}

The great flexibility of the hazard function $\lambda(t)$  of the continuous Weibull distribution and its ability to model various ageing mechanisms has made its success within the engineering community. 
Moreover, the parameters ($\eta$, $\beta$) of this model have a clearly understandable technical meaning. The first one is the quantile of the lifetime corresponding to a survival probability of approximately $1/3$ (actually 0.37) and the 
latter embodies (independently of the value of $\eta$) the nature of the ageing of the corresponding component or system: (i) rejuvenation if $\beta<1$, (ii) constant hazard rate if $\beta=1$ (exponential model), (iii) decelerated ageing if $\beta \in ]1,2[$, (iv) accelerated ageing if $\beta>2$. 
The transposition of these properties to the W1 distribution are investigated in the following. 
From the expression of $\lambda(n)$ (cf. Equations~\ref{eq:SurvivalAndHazardWeibull1}) it can be shown that:
\begin{itemize}
\item For $\beta=1$, the hazard function is constant. A trivial calculation gives: $\lambda(n)=1-\exp (-1/ \eta)$.
\item For $\beta>1$, the hazard function is an increasing function of $n$, as testified by  the argument of the exponential in the expression of $\lambda(n)$, for $n\geq 2$:
\begin{equation}\label{eq:ArgExpHazardWeibull1}
\left( \dfrac{n-1}{\eta} \right)^\beta - \left( \dfrac{n}{\eta}\right)^\beta .
\end{equation}
This function of $n$ is decreasing for $\beta>1$, thus $\lambda(n)$ is increasing.
\item for $\beta<1$, the hazard rate is a decreasing function of $n$, which can be shown following the same reasoning about the monotonicity of the function~(\ref{eq:ArgExpHazardWeibull1}) above, which is increasing if $\beta<1$.
\end{itemize}

The following proposition {\it (proven in Appendix)} states that decelerated aging can be diagnosed for the W1 model as for the usual Weibull model, by checking if  $\beta \in ]1, 2]$.
\begin{proposition}\label{prop-concavity}
For $\beta \in ]1, 2]$,  $\lambda(n)$ is a concave function of $n$.  Therefore the model is relevant to model decelerated ageing. 
\end{proposition}

For $\beta >2$ no analytical result about $\lambda(n)$  was obtained. However, as also highlighted in \cite{Xie2002}, following the classical definition of $\lambda(n)$ as a conditional probability, it is obvious that this function cannot be strictly convex as it must tend to $1$ as $n \rightarrow \infty$, which is not the case for the Weibull continuous model.



It was found empirically, by studying the convexity of $\lambda(n)$ for $(\eta,\beta) \in [1,1000] \times [2.1,20]$, that for a given $\beta$, a value $\eta_0$ of $\eta$ exists, so that for each $\eta<\eta_0$, $\lambda(n)$ is strictly concave and for each $\eta>\eta_0$, $\lambda(n)$ is initially convex, then concave, presenting thus an inflection point. The main lesson learnt by this empirical study is that, unlike the continuous Weibull distribution, the concavity of the hazard function does not depend on $\beta$ only, but also on $\eta$. As a conclusion, an interesting property of the Weibull model, particularly attractive for engineers, is actually lost when switching to W1. 

\begin{figure}[ht!]
\centering
\includegraphics[width=0.5\textwidth]{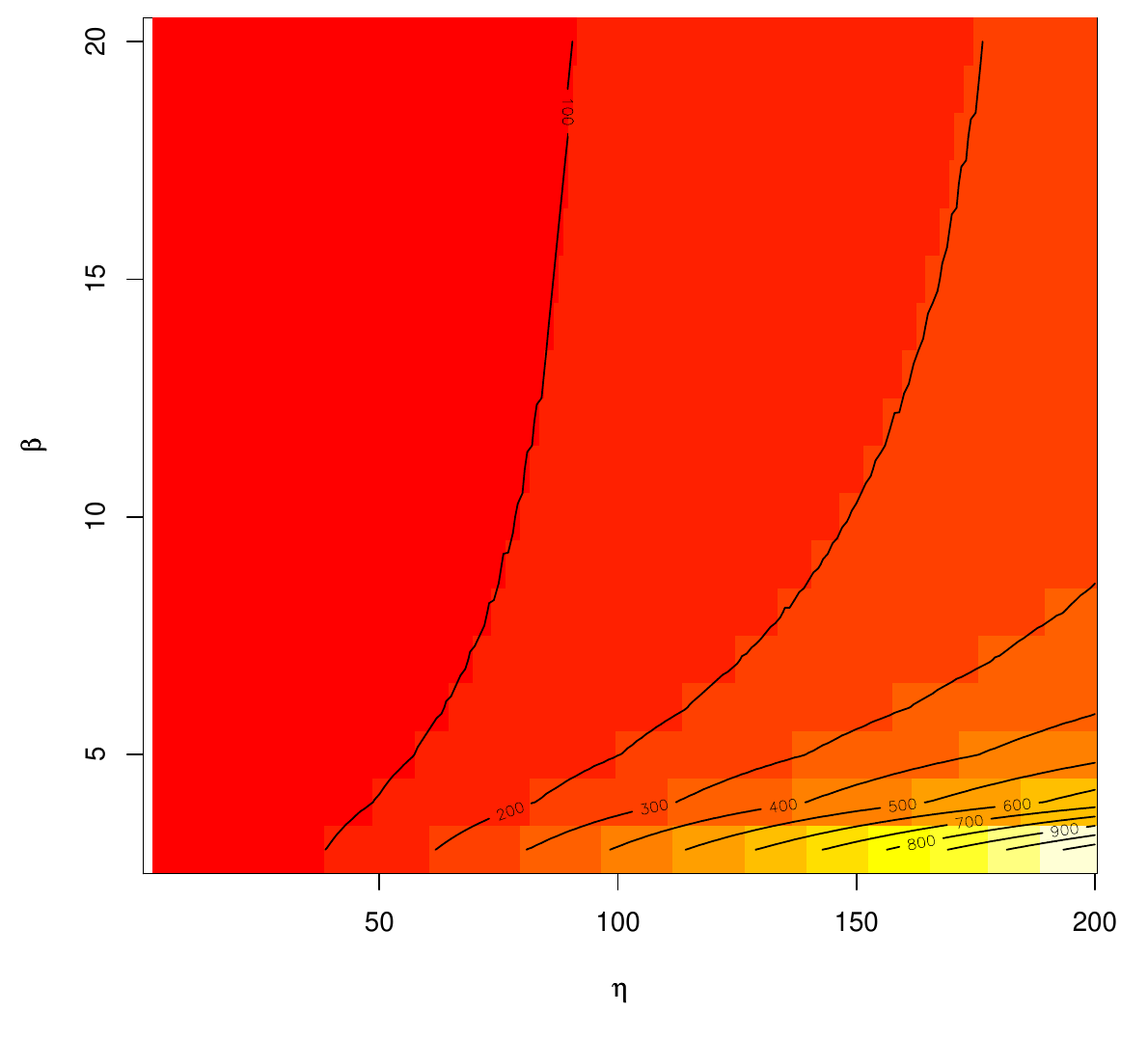}
\caption{\label{fig:InflexionPointsWeibull1}\small Weibull-1 model. Values of the solicitation number $n$ corresponding to the inflection points of the hazard function $\lambda(n)$ as a function of $\eta$ and $\beta$.} 
\end{figure}

Figure (\ref{fig:InflexionPointsWeibull1}) shows the value of $n$ corresponding to the inflection points, found by means of the empirical study described above. In practice, the presence of an inflection point is usually not a serious issue if the corresponding value of $n$ is a quantile corresponding to a very low survival probability. In that case, it can be concluded that for the set of values of $n$ interesting for practical purposes, $\lambda(n)$ is convex and can only embody phenomena of accelerated ageing. 

\begin{figure}[ht!]
\centering
\includegraphics[width=14cm,height=9cm]{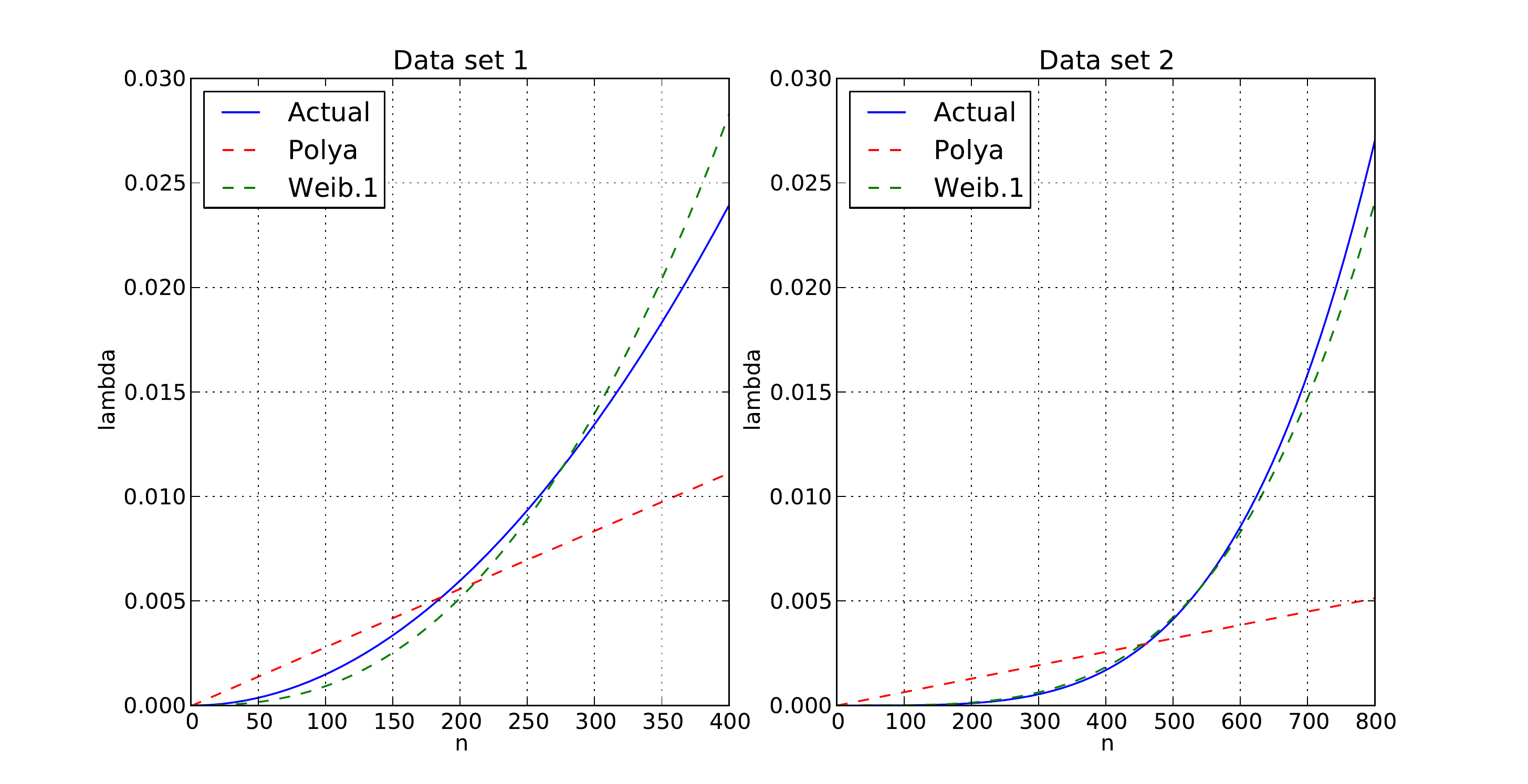}
\caption{\label{fig:SimulatedHazardAndData}\small Actual (blue curves) hazard functions and estimated hazard functions according to Weibull-1 and IPD models, the parameters of which have been estimated by maximizing the likelihood of random data sets of size $100$.} 
\end{figure}

The superiority of the W1 model over the IPD model, in terms of flexibility, can be tested by seeing how these two distributions are able to reproduce two known hazard functions from simulated samples. More precisely, starting from two known hazard functions, shown in Figure~\ref{fig:SimulatedHazardAndData} (blue curves), two samples of $100$ uncensored discrete lifetimes  were randomly generated. Then, using the two generated data sets, the Maximum Likelihood method was used to estimate the parameters of the IPD and W1 models (cf. Appendix and ref. \cite{Bracquemond2003}). Then the estimated hazard functions were plotted on Figure ~\ref{fig:SimulatedHazardAndData} (dotted curves). The relevance of this method to assess the quality of the adjustment in the reliability context is defended in \cite{Bracquemond2003}. Not surprisingly, in presence of convex hazard functions (i.e. accelerating ageing) the performance of the IPD model is poor. Intuitively, as IPD can only have concave hazard functions, the best approximation it can give of a convex $\lambda$ is a linear function. Instead, the flexible W1 model returns a quite fair approximation of $\lambda$. 


\section{Approximating W1 by the continuous Weibull model}\label{sec:Weibullcont}

\subsection{Formal results}\label{sec:formal}

Face to the issues sketched hereinbefore, it may appear practical for the engineer to use the usual (continuous) Weibull distribution for the assessment of main outcomes of interest in a reliability study. If seen as a possible "continuous approximation" of the Weibull-1 distribution, its computational treatment (e.g. parameter estimation, sampling) is well known and familiar to the practitioner. Consider the two random variables previously evoked in Section~\ref{sec:Weibull1Intro}:
\begin{equation}\label{eq:WAndW1Definition}
N \sim \text{W}_1(\eta,\beta) ~~~\text{and}~~~T \sim \text{W}(\eta,\beta).
\end{equation}
These two distributions have the same parameters but the first is discrete and the second is continuous. The closeness of both models appears  first in the proximity between both MTTF (noted ${\mathbb{E}}_{\text{W}_1} [ N|\eta,\beta ]$ and ${\mathbb{E}}_{\text{W}}[T|\eta,\beta]$, respectively), summarized in the next proposition, proven in Appendix. Especially, it highlights that the two MTTF's  become closer and closer as both quantities $\gg 1$. 

\begin{proposition}\label{prop1}
Given the two random variables $T$ and $N$ (defined by Equation~\ref{eq:WAndW1Definition}), the following inequality stands:
\begin{equation}\label{eq:EsperanceW}
 {\mathbb{E}}_{\text{W}}[T|\eta,\beta]   \leq \ {\mathbb{E}}_{\text{W}_1}[N|\eta,\beta] \ \leq   1+{\mathbb{E}}_{\text{W}}[T|\eta,\beta].
\nonumber
\end{equation} 
\end{proposition}

Pursuing the study of the closeness of $\text{W}_1(\eta,\beta)$ and $\text{W}(\eta,\beta)$, by definition, the survival functions of both models have the same mathematical form, namely they lead to the same value of the survival and the failure probability for a given $n$. Consequently, the expression of the quantile $n_q$ of probability $q$ is the same for $\text{W}_1(\eta,\beta)$ and $\text{W}(\eta,\beta)$: 
\begin{equation*}
n_q=t_q=\eta  \left[ -\log (1-q) \right]^{1/ \beta}.
\end{equation*}
Therefore, from an engineering viewpoint, the most interesting features and quantities of interest of both distributions (MTTF, quantiles, probability of failure) are similar when they have the same parameters.
Moreover, when estimating $(\eta,\beta)$ from actual industrial feedback data in presence of right-censored observations, the likelihoods of the two models tend also to be very close: the inference, thus, leads to very similar estimates for $\eta$ and $\beta$ for both models. Actually, it can be seen that the likelihoods of a given samples of discrete lifetime, according to  $\text{W}_1(\eta,\beta)$ and $\text{W}(\eta,\beta)$ respectively, are closer and closer as (i) the rate of right-censored data increases, and (ii) the (unknown) value of $\eta$ is high.

The proof of the first part of the assertion is trivial: any right-censored datum $n$ contributes to the likelihood by means of the value of the survival function $S(n| \eta,\beta)$, which has the same expression for both Weibull and Weibull-1 distributions.
As far as the second part of the proposition is concerned, denoting $f_{\text{W}}(\cdot)$  the density of $\text{W}(\eta,\beta)$ and $p_{\text{W}_1}(\cdot)$ the probability distribution of $\text{W}_1(\eta,\beta)$, the contribution of an uncensored observation $n$ to the likelihood of the two models is equal to $f_{\text{W}}(n)$ and $p_{\text{W}_1}(n)$, respectively. 

\begin{figure}[ht!]
\centering
\includegraphics[width=0.99\textwidth]{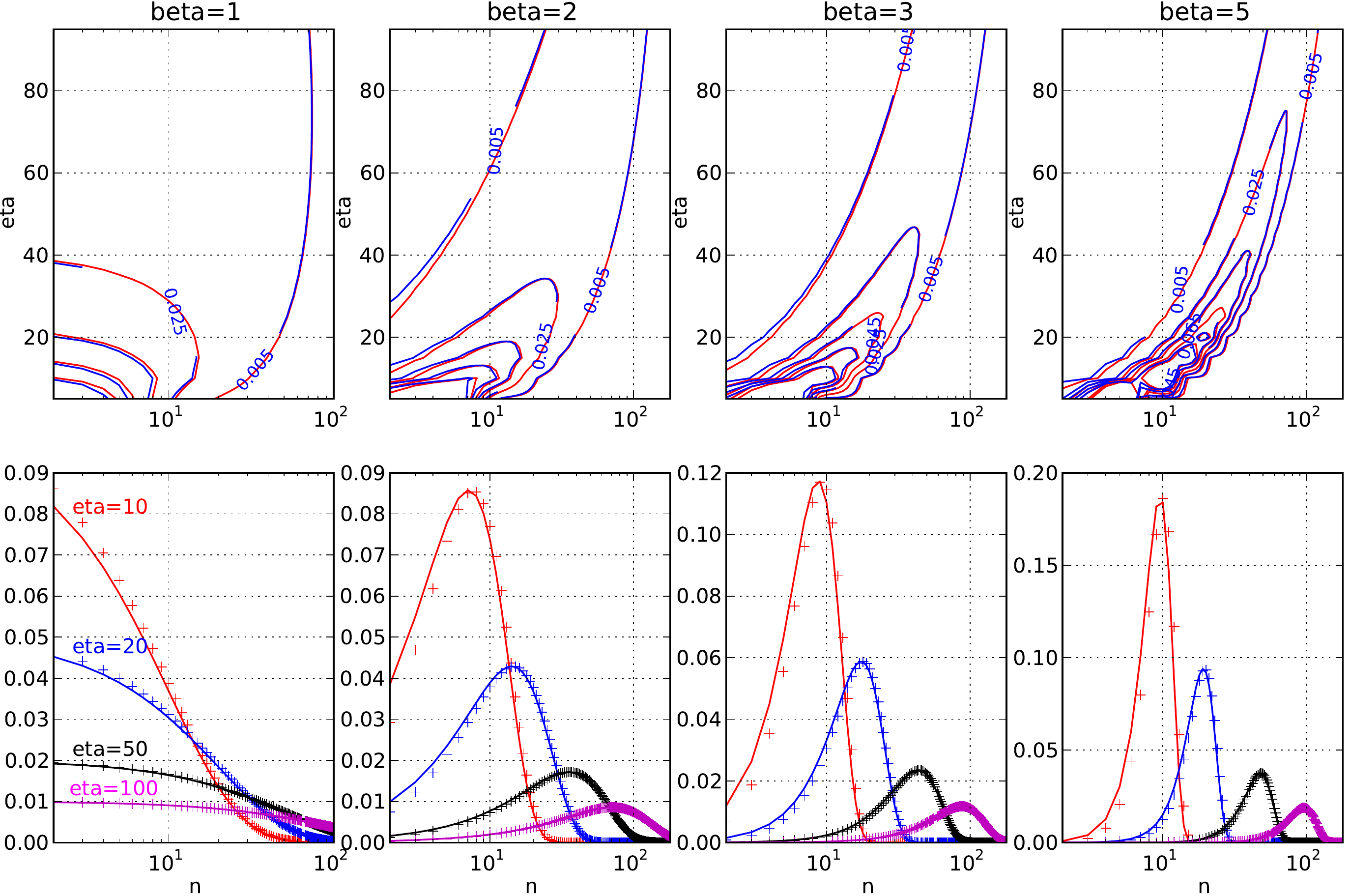}
\caption{\label{fig:W1vsW}\small Upper part: plot of the Weibull-1 (red lines) and Weibull distribution functions (blue lines) for fixed values of $\beta$ as a function of $n$ and $\eta$; one can see that isolines are very close to one another. Lower part: plot of the discrete Weibull-1 distribution (crosses) vs. the corresponding values of the continuous Weibull for given values of $\eta$ and $\beta$ (continuous lines), as functions of $n$; as one can see, the higher $\eta$ and $n$ are, the more the crosses tend to be superposed over the lines.}
\end{figure}

Since the survival functions $S(\cdot)$ have the same expression for $\text{W}_1(\eta,\beta)$ and $\text{W}(\eta,\beta)$, one may write \cite{Lai2013ASMBI}:
\begin{eqnarray}\label{eq:LinkDensityWpdfW1}
p_{\text{W}_1}(n)  & = & \mathbb{P}[N>n-1]- \mathbb{P}[N>n], \nonumber\\
                             & = & S(n-1)-S(n), \nonumber \\ 
                             & = & \int\limits_{n-1}^\infty	f_{\text{W}}(t)dt - \int\limits_{n}^\infty f_{\text{W}}(t)dt,\nonumber  \\
                             & = & \int\limits_{n-1}^n f_{\text{W}}(t)dt .
\end{eqnarray}
It is easy to provide the following bounds for the last integral in the right hand side of Equation~\ref{eq:LinkDensityWpdfW1}:
\begin{equation}\label{eq:LinkDensityWpdfW1bis}
\underset{t\in [n-1,n]}\min f_{\text{W}}(t) \leq p_{\text{W}_1}(n) \leq \underset{t\in [n-1,n]}\max f_{\text{W}}(t).
\end{equation}
Intuitively, the higher the values of $\eta$ and $n$, the closer the bounds in Equation \ref{eq:LinkDensityWpdfW1}, and consequently, the closer $p_{{\text{W}}_1}(n)$ and  $f_{\text{W}}(n)$. The graphics displayed on Figure~\ref{fig:W1vsW} confirm, empirically, this intuition.

More formally it is proven in Appendix the following result. 

\begin{proposition}\label{prop2}
For all $\beta\geq 1$:
\begin{eqnarray*}
\lim_{\eta \to \infty}  \sup_{t \in \mathbb R^+} |p_{{\text{W}}_1}(t|\beta,\eta) - f_\text{W}(t|\beta,\eta)| = 0. \\
\end{eqnarray*} 
\end{proposition}

In addition, note also that the convergence to $0$ of $|p_{{\text{W}}_1}(t|\beta,\eta) - f_\text{W}(t|\beta,\eta)|$ when $t \to\infty$ is trivial as both $p_{{\text{W}}_1}(t|\beta,\eta)$ and $f_\text{W}(t|\beta,\eta)$ tend to $0$.

In the following paragraph, these properties of approximation are exemplified and highlighted through an empirical study conducted using simulated datasets.

\subsection{Numerical experiments}\label{sec:W1EmpiricalStudy}\label{sec:experiments}


The previous results suggest that for practical industrial purposes (i.e. predicting probabilities of failure and MTTF) the W1 and Weibull models provide very similar outcomes when $\eta$ is high. High values of $\eta$ mean that the system under investigation is \textit{reliable} in the common-sense meaning, that is failures occur for \textit{high} values of $n$ (i.e. $\gg 1$). Moreover, industrial feedback datasets contain generally a number of censored data. In particular, in the specific industrial context considered here, the data are most of the time right-censored (and quite never left-censored) because failures are to be strictly avoided as they have a costly impact on availability of the overall production facility. \\

As shown previously, a set of lifetimes of a reliable system with a significant number of censored data leads to a very similar likelihood under the two hypotheses of Weibull-1 and Weibull model.
Hence, Maximum Likelihood estimations (MLE) of ($\eta,\beta$) for both models are expected to be very close.
To confirm these results, intensive numerical simulations were conducted, by generating datasets likely to be encountered in industrial practice and thus evaluating the MLE of ($\eta,\beta$) for Weibull-1 and Weibull model, noted $(\hat\eta_{\text{W}_1},\hat\beta_{\text{W}_1})$ and  $(\hat\eta_{\text{W}},\hat\beta_{\text{W}})$, respectively. \\

More precisely, for $(\eta,\beta) \in \{10, 50, 300, 500, 800, 1000\} \times \{0.5, 1, 1.5, 2, 2.5, 3, 5, 10 \}$, and for right-censored data rates of 0\%, 25\%, 50\% and 75\%, 5000 samples of sizes 50 and 100 were generated from the Weibull-1 distribution $\text{W}_1(\eta,\beta)$. Based on these data, the MLE $(\hat\eta_{\text{W}_1},\hat\beta_{\text{W}_1})$ and  $(\hat\eta_{\text{W}},\hat\beta_{\text{W}})$ were evaluated, as well as the relative errors concerning the estimations of $(\eta,\beta)$:
\begin{equation*}
\dfrac{\eta - \hat\eta_{\text{W}_1}}{\eta}, ~~ \dfrac{\beta - \hat\beta_{\text{W}_1}}{\beta}, ~~
\dfrac{\eta - \hat\eta_{\text{W}}}{\eta}, ~~ \dfrac{\beta - \hat\beta_{\text{W}}}{\beta}
\end{equation*}
and the relative errors of plug-in estimators of the following quantities of interest: hazard rates corresponding to the quantiles of probabilities $(0.5, 0.75, 0.90, 0.99)$ of the original distribution, MTTF and quantiles.
As data were generated from W1 distributions, the estimators $(\hat\eta_{\text{W}_1},\hat\beta_{\text{W}_1})$ are expected to be closer to the actual values of $(\eta,\beta)$ than $(\hat\eta_{\text{W}},\hat\beta_{\text{W}})$. \\ 

\begin{figure}[ht!]
\centering
\includegraphics[width=0.99\textwidth]{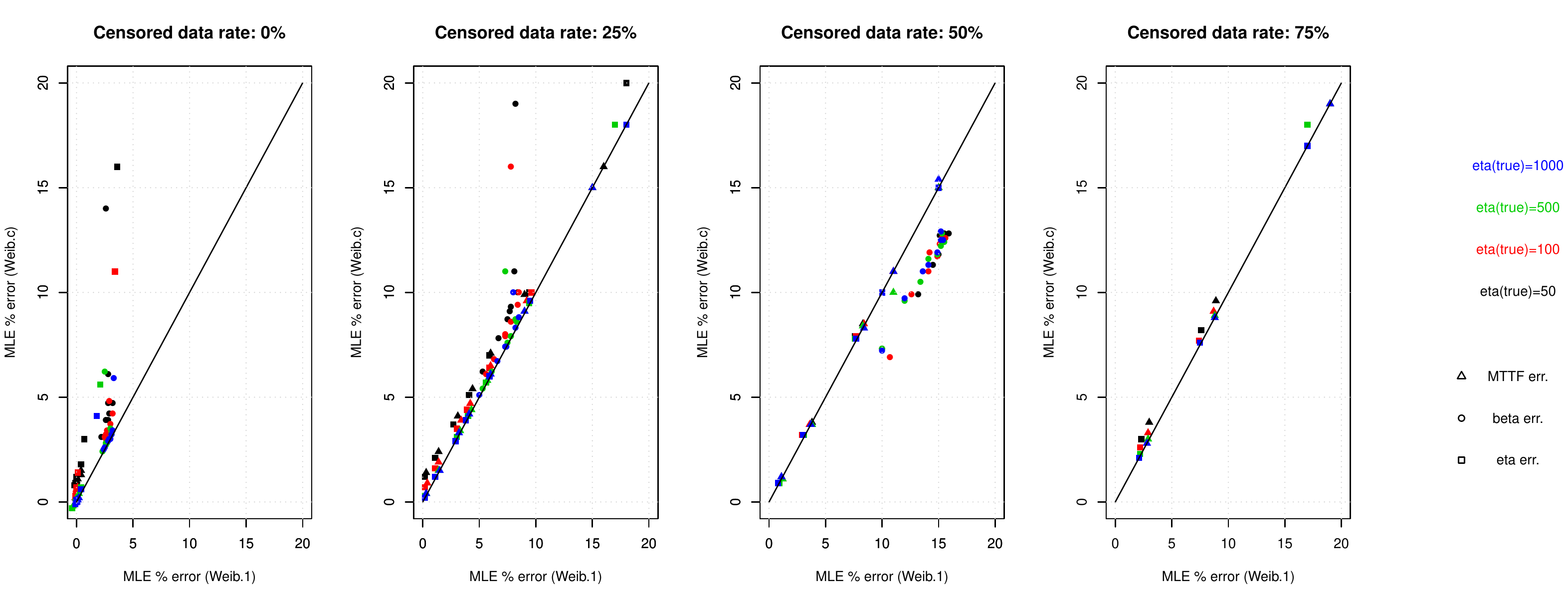}
\caption{\label{fig:EstimationEtaBeta}\small Relative errors of the MLE of $\eta$ (circles) and $\beta$ (squares) and of the ML plug-in estimator of the MTTF (triangles), obtained from data simulated from the Weibull-1 model under the hypothesis of different censoring rates from 0\% to 75\%. The estimation is carried using the Weibull-1 and the continuous Weibull model assumption. The colors correspond to different values of the \textit{true} $\eta$.}
\end{figure}

On Figure~\ref{fig:EstimationEtaBeta} are displayed the main results of this empirical study. The mean ML estimation error under  the W1 model assumption (x-axis) is plotted against the error under the continuous Weibull model assumption. It can be seen that the points of the scatterplot are quite close to the first bisector, showing that using the W1 model yields no significant improvement with respect to the continuous approximation.

\section{Inference and ageing diagnostic from actual feedback data}\label{sec:real}

Two EDF datasets coming from actual industrial feedback are considered in this last technical section. Even though this analysis is proposed for exemplary purposes only, these data are nevertheless representative of those reliability engineers cope with in numerous fields. On Table~\ref{tab:DataDiscreteREX} are summarized the main features of the examined datasets, that allow to find out two important characteristics. First, most of the data are right-censored lifetimes: the censoring rates are equal to 96\% and 81\%, respectively. Second, the components under investigation are \textit{reliable}, in the sense that failures are expected to occur after a (relatively) high number of solicitations: the empirical means of the (highly censored) observed data are 63.8 and 334.5 respectively. 

\small
\begin{table}[!h]
\centering
\begin{tabular}{|c|c|c|}
\hline 
 & {\small Sample 1 (Aux. power device } & {\small Sample 2 }\tabularnewline
 & {\small linings) \cite{Clarotti1997}) } & {\small $ $ }\tabularnewline
\hline 
\hline 
{\small Data size } & {\small 497 } & {\small 48 }\tabularnewline
\hline 
{\small Sum of observed data } & {\small 31715 } & {\small 16058 }\tabularnewline
\hline 
{\small Observed failures } & {\small 18 } & {\small 9 }\tabularnewline
\hline 
{\small Number of right-censors } & {\small 479 } & {\small 39 }\tabularnewline
\hline \hline
\multicolumn{1}{|c}{\textbf{\small Parameters estimation}{\small{} }} & \multicolumn{1}{c}{{\small $ $ }} & {\small $ $ }\tabularnewline
\hline 
{\small Inv. P\'{o}lya } & {\small $\begin{array}{c}
\hat{\alpha}=7.037\cdot10^{-12}\\
\hat{\zeta}=1.349\cdot10^{-5}
\end{array}$ } & {\small $\begin{array}{c}
\hat{\alpha}=5.601\cdot10^{-4}\\
\hat{\zeta}=1.774\cdot10^{-19}
\end{array}$ }\tabularnewline
\hline 
{\small Weibull-1 } & {\small $\begin{array}{c}
\hat{\eta}=306.814\\
\hat{\beta}=2.320
\end{array}$} & {\small $\begin{array}{c}
\hat{\eta}=1530.139\\
\hat{\beta}=1.122
\end{array}$}\tabularnewline
\hline 
{\small Weibull } & {\small $\begin{array}{c}
\hat{\eta}=320.580\\
\hat{\beta}=2.320
\end{array}$} & {\small $\begin{array}{c}
\hat{\eta}=1510.250\\
\hat{\beta}=1.124
\end{array}$}\tabularnewline
\hline 
\end{tabular}
\caption{\label{tab:DataDiscreteREX}\small Example of analysis of data set coming from actual industrial feedback. Upper part: data summary. Lower part: Maximum Likelihood estimators of the parameters of Inverse P\'{o}lya, Weibull-1 and Weibull distribution.}
\end{table}
\normalsize

\begin{figure}[htp!]
\centering
\includegraphics[width=0.5\textwidth]{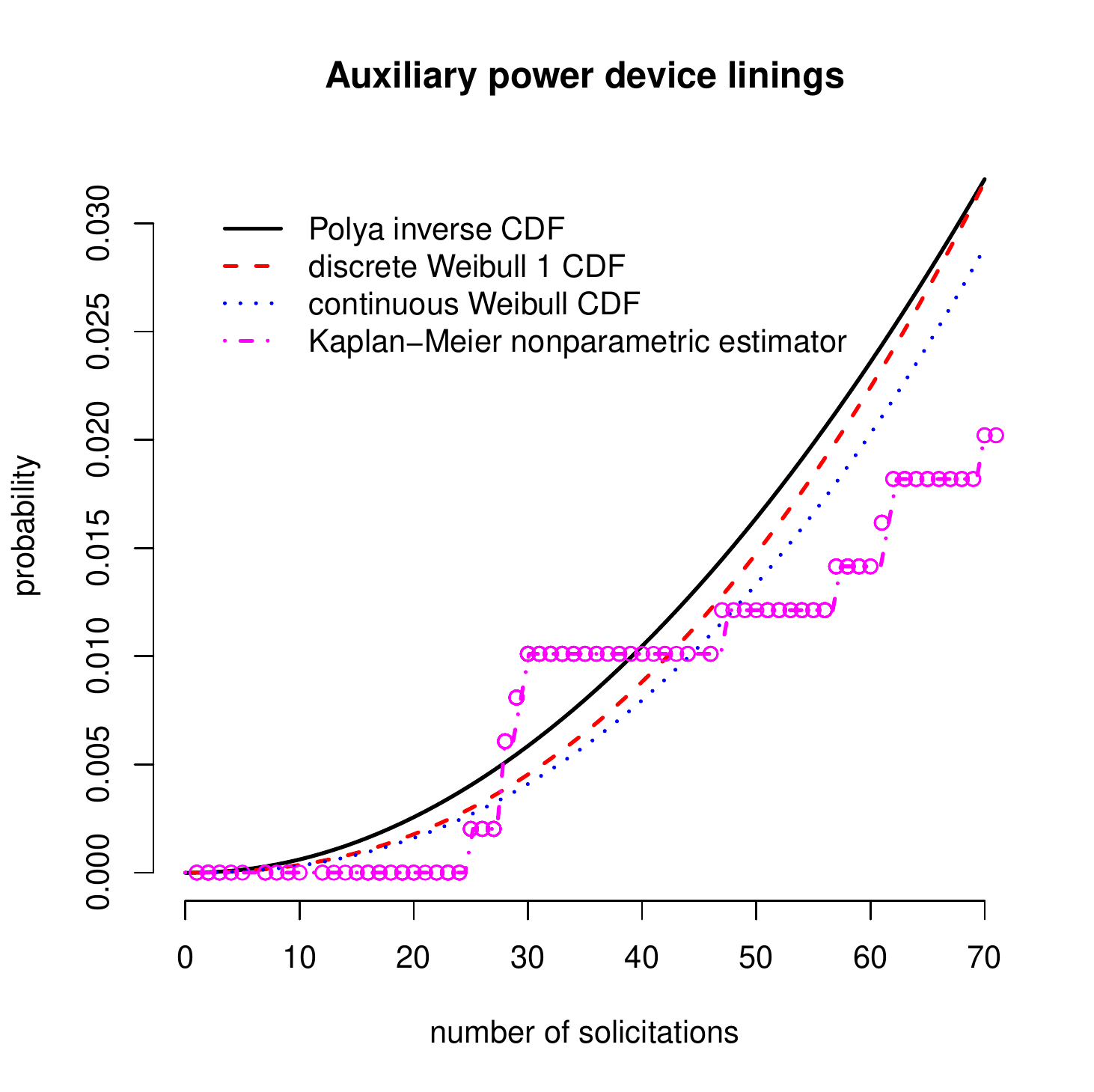}
\caption{\label{fig:AuxiliaryLiningsZoom}\small Data Sample 1: Cumulative distribution functions from ML estimations of Inverse P\'{o}lya, Weibull and Weibull-1 model and non-parametric Kaplan-Meier estimator.}
\end{figure}

\begin{figure}[htp!]
\centering
\includegraphics[width=0.5\textwidth]{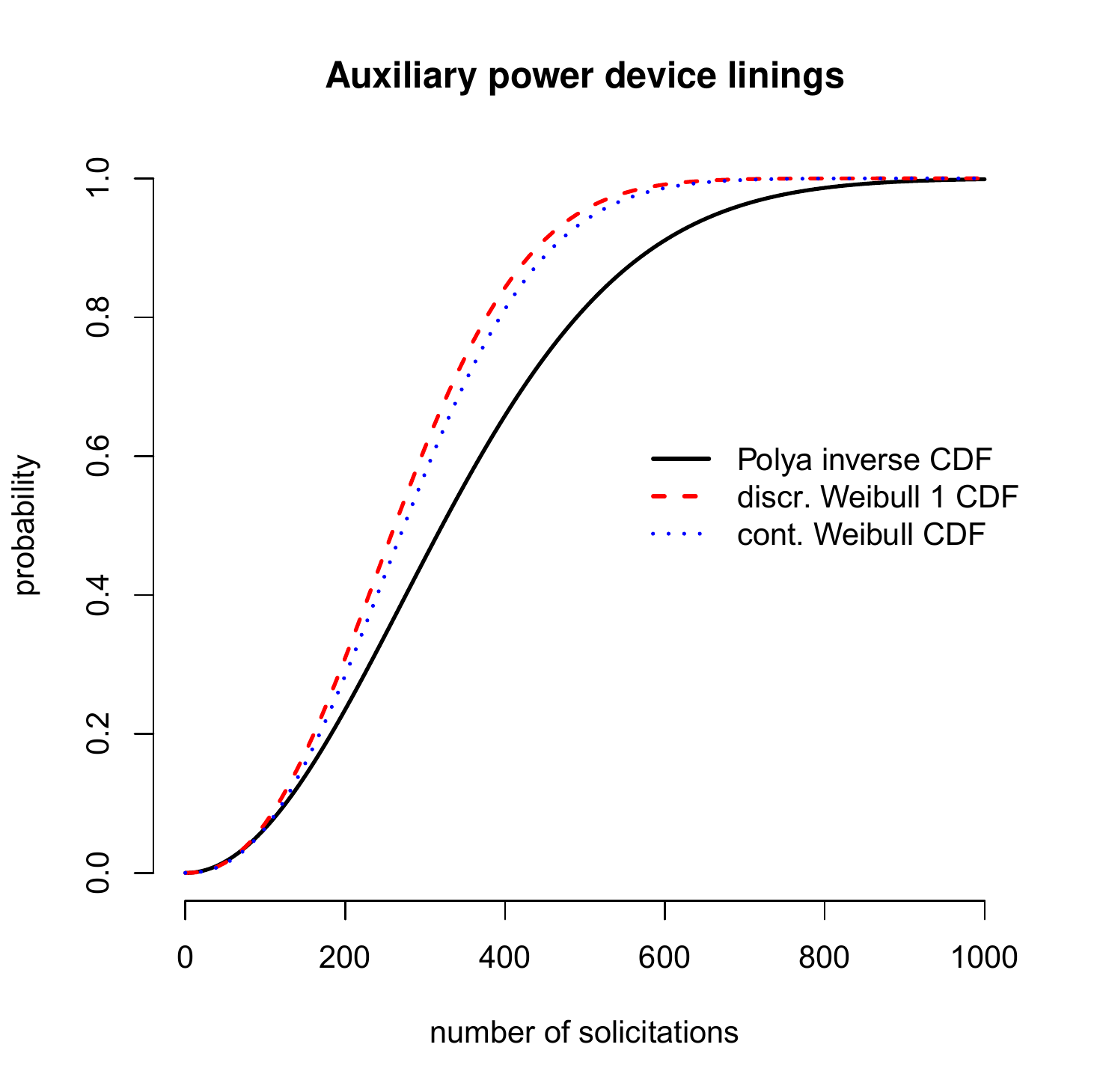}
\caption{\label{fig:AuxiliaryLinings}\small Data Sample 1: Cumulative distribution functions from ML estimations of Inverse P\'{o}lya, Weibull and Weibull-1 model. The range of the number of solicitation is here extended beyond the maximum of the observed sample to show the predictive properties of the model for high values of $n$.}
\end{figure}

In the same table are also shown the ML estimators of the parameters of the Inverse P\'{o}lya (IPD), W1 and (continuous) Weibull models.
Regarding the first dataset, the estimated parameters of both W1 and Weibull models (which have very similar values) suggest  accelerated ageing. The extremely high (and hardly understandable by a technical viewpoint) value of ratio $\hat\zeta / \hat\alpha$ (order of magnitude: $10^6$) highlights a poor modelling performance of the IPD model.
On Figure~\ref{fig:AuxiliaryLiningsZoom} are displayed the cumulative distribution functions (CDF) of the three estimated distributions as well as the usual Kaplan-Meier estimator. In spite of the issues evoked hereinbefore, the prediction properties of the three models (in terms of failure probabilities) are quite equivalent within the range of observed data. Yet, as shown in Figure~\ref{fig:AuxiliaryLinings}, the predictions given by IPD for higher values of $n$ are more optimistic and less conservative, in the sense IPD provides lower values of the CDF (i.e. higher values of the reliability function) than the ones given by Weibull and Weibull-1, the CDF's of which are practically indistinguishable.

\begin{figure}[htp!]
\centering
\includegraphics[width=0.5\textwidth]{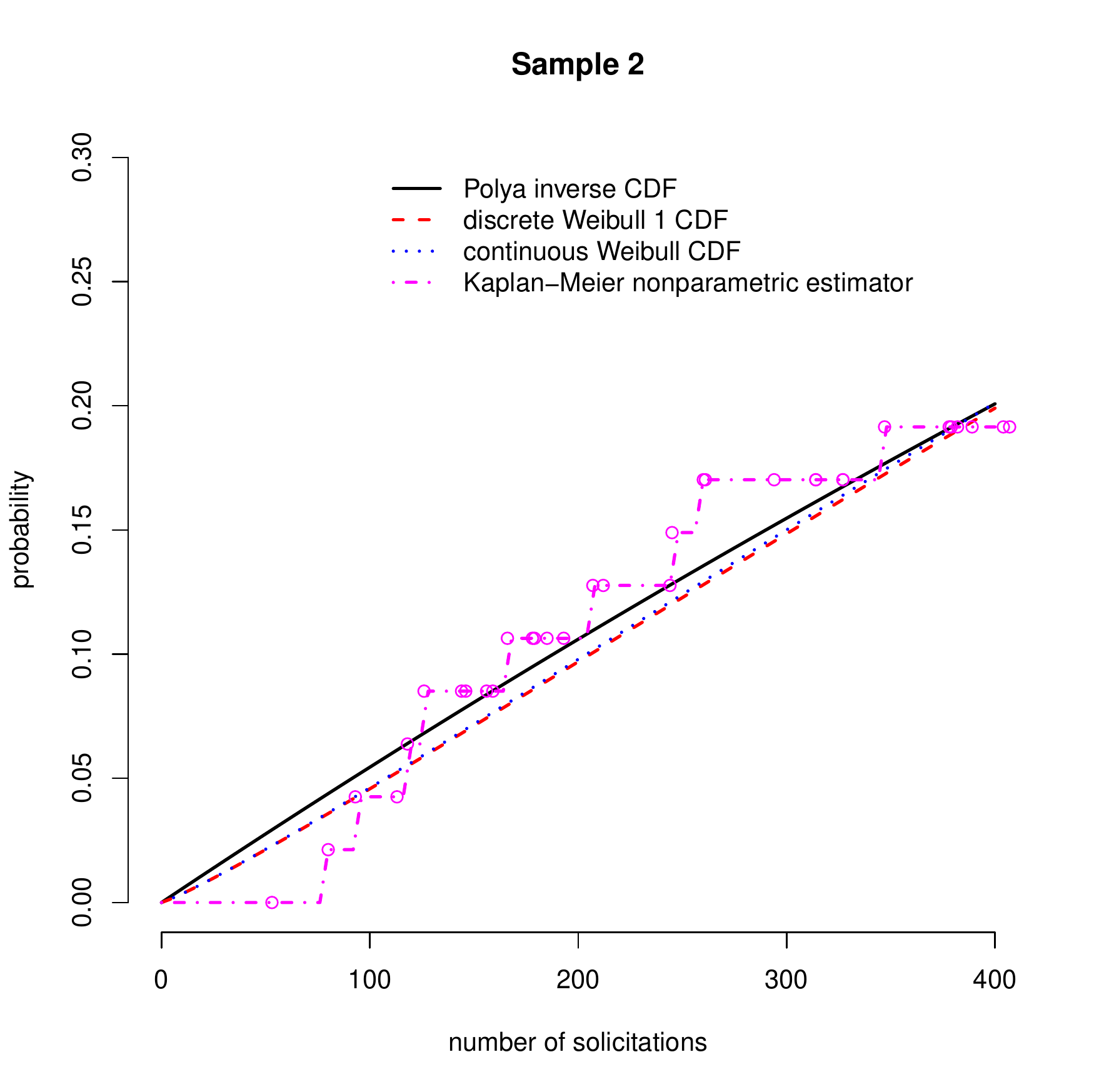}
\caption{\label{fig:Sample2}\small Data Sample 2: Cumulative distribution functions from ML estimations of Inverse P\'{o}lya, Weibull and Weibull-1 model and non-parametric Kaplan-Meier estimator.}
\end{figure}

As far as the Sample 2 is concerned (cf. Figure~\ref{fig:Sample2}), the components do not show a significant ageing (the Weibull shape parameter is close to 1). The three probabilistic models return a very similar prediction in terms of CDF (and reliability function). 

As a conclusion, these exemplary analyses confirm the conclusions presented in the previous sections, by means of theoretical and empirical considerations: for engineering purposes, the continuous Weibull model is a fairly good alternative to the discrete model (Weibull-1) investigated in the framework of the present study. 

\paragraph{Remark.}  {\it Although the data come from real surveys, the study conducted in this section is given for exemplary purposes only and neither results nor methodology must be extrapolated to make any general conclusion about the reliability of EDF industrial components or EDF risk assessment policies.}


\section{Discussion}
The study shown hereinbefore has highlighted some weaknesses of both inverse P\'{o}lya (IPD) and Weibull-1 (W1) distributions as discrete models for lifetime of industrial components.
The IPD model carries the implicit hypothesis of decelerating ageing. This can be an issue as this  assumption can sometimes be hardly justified a priori in industrial studies. Nonetheless, conducting wise preventive maintenance strategies
can ensure that the perceived ageing of a component remains low and decreases through time (for instance by benefiting of technological improvements during the maintenance operations). In such cases, the IPD model appears a very practical and intuitive model for the engineer. 

As far as the Weibull-1 model is concerned, it has been shown that the popular interpretation of the shape and scale parameters of the Weibull distribution is no longer (fully) valid for its discrete version.
In particular, the type of ageing that the model can embody does not depend on the shape parameter $\beta$ only but also on the scale parameter $\eta$. 

Moreover, for practical purposes, the W1 model and the Weibull model are very close. In practice, starting from the same (discrete) data the Maximum Likelihood estimation of the parameters ($\eta,\beta$) computed under the hypotheses of W1 and Weibull models leads to close numerical results, as far as the usual reliabilty quantities of interest are concerned. This closeness increases with the value of $\eta$ (i.e. the piece of equipment under investigation is reliable, in the sense that it normally fails after a significantly high number of solicitations) and the rate of censored data. This characterizes exactly lifetime data collected within industrial processes where  fortuitous  failures can have a great impact on the availability of the production facilities and lead to high unexpected costs (as in the particular context of EDF). 
Thus, the practical impact of the use of W1 model for improving reliability analyses based on feedback data is quite low, and our advice is rather to use a continuous Weibull distribution in the situations uncovered by the IPD model. 

Nonetheless, the easily-interpretable features of the Inverse P\'{o}lya distribution could remain valuable in practice if the phenomenon of decelerating ageing could be discarded. It is likely that adding a supplementary hypothesis of the following nature could improve the versatility of the model: namely, the number $z$ of balls added at solicitation $n$ should follow an increasing  pattern in function of $n$ rather than remaining constant. Defining and comparing several patterns, from both analytical and computational viewpoints, should be a keypoint of future studies aiming at preserving the interest of IPD in reliability analysis. 
\\



\paragraph*{Acknowledgements.} Most of this work has been realized when C\^ome Roero was research
fellow at INRIA - Paris Sud, under the supervision of Prof. Gilles Celeux, to whom the authors
owe particular debts of gratitude for his contribution and his ideas.
The authors gratefully thank Dr. Merlin Keller (EDF R\&D) for numerous fruitful discussions and for having formally provided the proof of Proposition~(\ref{prop2}).

\newpage
\bibliographystyle{plain}
\bibliography{biblioDiscreteModels}


\newpage
\section*{Appendix}

\paragraph{Sampling from the inverse P\'olya distribution}\label{simu.polya}
Simulating a $n-$sample from the inverse P\'olya distribution can be numerically done using the following algorithm: 
\texttt{
\begin{itemize}
\item Set $n_1=n$. 
\item Step $k\rightarrow k+1$, for $k\geq 1$:
\begin{enumerate}
\item simulate a $n_k-$sample $x^{(k)}_{1},x^{(k)}_{n_k}$ from the Bernoulli distribution with parameter
\begin{eqnarray*}
\alpha_k & = & \frac{\alpha + (k-1)\zeta}{1+(k-1)\zeta};
\end{eqnarray*}
\item compute ${\displaystyle n_{k+1} = \sum_{i=1}^{n_k}
\left(1-x^{(k)}_i\right)}$ ; \item add to the sample the value  $k$ replicated $\sum_{i=1}^{n_k}x^{(k)}_i$ times;
\item stop if $n_{k+1}=0$. 
\end{enumerate}
\end{itemize}
}

\paragraph{Maximum likelihood estimation of the inverse P\'olya distribution}\label{mle.polya}
Assume that among the available data ${\bf D}$  there are $s$ survival data and $r$ failure observations. Besides, assume that 
$k_i$ components have survived until the $n_i-$th solicitation, for $i=1,\ldots,s$, and 
that the $r$ components have broken down after $n_{s+1},\ldots,n_{s+r}$ solicitations, respectively. Denoting
\begin{eqnarray*}
\alpha & = & \sum\limits_{i=1}^s k_i n_i + \sum\limits_{j=s+1}^{s+r} n_j - r,
\end{eqnarray*}
the likelihood of observed data ${\bf D}$  is written as
\begin{eqnarray}
{\cal{L}}({\bf D}|\alpha,\zeta) & = & \frac{\left(1-\alpha\right)^{\alpha} \prod\limits_{i=s+1}^{s+r} \left(\alpha + (n_i-1)\zeta\right)}{\left[\prod\limits_{j=1}^s \left(\prod\limits_{k=1}^{n_j-1} (1+j\zeta)\right)^{k_j}\right]\left[\prod\limits_{l=s+1}^{s+r} \left(\prod\limits_{p=1}^{n_l-1} (1+p\zeta)\right)\right]}, \label{vraispolya}\\
\nonumber & = & \frac{\left(1-\alpha\right)^{\alpha} \Gamma^{\mu}(1/\zeta)  \prod\limits_{i=s+1}^{s+r} \left(\alpha + (n_i-1)\zeta\right)}{\zeta^{\alpha+r} \left[\prod\limits_{j=1}^s \Gamma^{k_j}(n_j+1/\zeta)\right]\left[\prod\limits_{k=s+1}^{s+r} \Gamma(n_k+1/\zeta)\right]} \ \ \text{with $\mu  =   \sum\limits_{i=1}^s k_i + r$.} \label{lik1}
\end{eqnarray}
The parameters $(\alpha,\zeta)$ are assumed to take their values in spaces $\Theta\in[0,1]$ and
$\Delta\in[0,\infty[$. A Newton-Raphson descent algorithm for assessing the maximum likelihood estimator (MLE) can be carried out, based on the following rationale. Denoting 
$\ell(\alpha,\zeta)=\log {\cal{L}}({\bf D}|\alpha,\zeta)$, the equation $\frac{\partial \ell}{\partial \alpha} = 0$ implies that
\begin{eqnarray}
\sum\limits_{i=s+1}^{s+r} \frac{1-\alpha}{\alpha + (n_i-1)\zeta}
& = & \alpha,\label{vrais1}
\end{eqnarray}
and since $\partial^2 \ell/{\partial \alpha^2}<0$, (\ref{vrais1}) has a unique solution maximizing (\ref{lik1}) given $\zeta$. Furthermore, $\frac{\partial \ell}{\partial \zeta} = 0$ implies that
\begin{eqnarray}
\sum\limits_{i=s+1}^{s+r} \frac{\zeta^2}{\alpha + (n_i-1)\zeta}
- (\alpha+r)\zeta - \sum\limits_{j=1}^{s+r+1}
k'_j\Psi(n'_j+1/\zeta) & = & 0 \label{vrais2}
\end{eqnarray}
where $k'_j=k_j$ for $j=1,\ldots,s$, $k'_j=1$ for
$j=s+1,\ldots,s+r$ and $k'_{s+r+1}=\mu$. Besides, $n'_j=n_j$
for $j=1,\ldots,s+r$ and $n'_{s+r+1}=0$. Denoting $\Psi$  the digamma function, a descent algorithm can solve 
(\ref{vrais1}) and
(\ref{vrais2}), following the iterative scheme
\begin{eqnarray*}
\left(\begin{array}{l} \alpha_{n+1} \\
\zeta_{n+1}\end{array}\right) & = & \left(\begin{array}{l}
\alpha_{n} \\ \zeta_{n}\end{array}\right) - \lambda_n
F_n\nabla^{-1}_n
\end{eqnarray*}
where $\lambda_n$ is an adaptive step, $F_n=F(\alpha_n,\zeta_n)$,
$\nabla_n=\nabla(\alpha_n,\zeta_n)$,
\begin{eqnarray*}
F(\alpha,\zeta) & = & \left(\begin{array}{l} \frac{\alpha}{1-\alpha} - \sum\limits_{i=s+1}^{s+r} \frac{1}{\alpha + (n_i-1)\zeta} \\
\sum\limits_{i=s+1}^{s+r} \frac{\zeta^2}{\alpha + (n_i-1)\zeta}
-{(\alpha+r)}{\zeta} - \sum\limits_{j=1}^{s+r+1}
k'_j\Psi(n'_j+1/\zeta)
\end{array}\right)
\end{eqnarray*}
and
\begin{eqnarray*}
\nabla(\alpha,\zeta) & = & \left(\begin{array}{llll}
\frac{\alpha}{(1-\alpha)^2} + \sum\limits_{i=s+1}^{s+r} \frac{1}{\left(\alpha + (n_i-1)\zeta\right)^2} & & \sum\limits_{i=s+1}^{s+r} \frac{n_i-1}{\left(\alpha + (n_i-1)\zeta\right)^2} \\
-\sum\limits_{i=s+1}^{s+r} \frac{\zeta^2}{\left(\alpha +
(n_i-1)\zeta\right)^2} && \sum\limits_{i=s+1}^{s+r}
\frac{2\alpha\zeta}{\left(\alpha + (n_i-1)\zeta\right)^3} +
{(\alpha+r)} + \sum\limits_{j=1}^{s+r+1}
\frac{k'_j}{\zeta^2}\Psi'(n'_j+1/\zeta)
\end{array}\right).
\end{eqnarray*}
In practice, the step $\lambda_n$ can be calibrated in function of the bounds of the parametric space (hence $\alpha_{n}\in[0,1]$ $\forall n$) and the possible lack of inversibility of $\nabla_n$. It is recommended to initialize the method by using a crude likelihood maximization over a grid of $\Theta\times\Delta$. 

\paragraph{Proof of Proposition \ref{prop1}: bounds for the MTTF of Weibull-1 distribution.}
Let write the expression of the mean of $\text{W}_1(\eta,\beta)$, taking into account  expression~(\ref{eq:LinkDensityWpdfW1}), showing the link between the density $f_{\text{W}}(\cdot)$  and the discrete pdf $p_{\text{W}_1}(\cdot)$:
\begin{equation*}
\mathbb{E}_{\text{W}_1}(N)=\sum\limits_{i=1}^{\infty} i \, p_{\text{W}_1}(i) =\sum\limits_{i=1}^{\infty} \, \int\limits_{i-1}^i i \, f_{\text{W}}(t)dt
= \sum\limits_{i=1}^{\infty} \, \int\limits_{i-1}^i ([t]+1) \, f_{\text{W}}(t)dt = \int\limits_{0}^\infty ([t]+1) \, f_{\text{W}}(t)dt,
\end{equation*}
$[t]$ being the floor function of the random variable $t\sim\text{W}(\eta,\beta)$. As $t-1<[t]\leq t+1 \Rightarrow tf_{\text{W}}(t)<([t]+1)\leq(t+1)f_{\text{W}}(t)$ by integrating this inequality over $t$ one concludes that:
\begin{equation*}
 \int\limits_{0}^\infty t \, f_{\text{W}}(t)dt < \mathbb{E}_{\text{W}_1}(N) \leq \int\limits_{0}^\infty t \, f_{\text{W}}(t)dt +  \int\limits_{0}^\infty 1 \, f_{\text{W}}(t)dt,
\end{equation*}
i.e. $ \mathbb{E}_{\text{W}}(T) <  \mathbb{E}_{\text{W}_1}(N) \leq  \mathbb{E}_{\text{W}}(T)+1$.

\paragraph{Proof of Proposition \ref{prop-concavity}: Concavity of Weibull-1 hazard function when $1<\beta\leq 2$.} 
First notice that for $n\geq 3$:
\begin{equation*}
\lambda''(n)=\lambda(n)-2\lambda(n-1)+\lambda(n-2) = \exp \left[ - \left( \dfrac{n-1}{\eta} \right) ^\beta + \left( \dfrac{n-2}{\eta} \right) ^\beta \right] \cdot A(n)\\
\end{equation*}
with:
\begin{equation*}
A(n)=2-\exp \left[ -\left( \dfrac{n}{\eta} \right) ^\beta +2 \left( \dfrac{n-1}{\eta} \right) ^\beta - \left( \dfrac{n-2}{\eta} \right) ^\beta \right] -\exp \left[ \left(  \dfrac{n-1}{\eta} \right) ^\beta - 2 \left( \dfrac{n-2}{\eta} \right) ^\beta + \left( \dfrac{n-3}{\eta} \right) ^\beta \right].
\end{equation*}
Consider now the function $G(x)=2-\exp(-x)-\exp(x)$. For $x>0$,  its first derivative $G'(x)=\exp(-x)-\exp(x)$ is negative; $G(x)$ is strictly decreasing and as $G(0)=0$, it also negative for $x>0$.
For $\beta>1$, $G \left[  \left(  \dfrac{n-1}{\eta} \right) ^\beta - 2 \left( \dfrac{n-2}{\eta} \right) ^\beta + \left( \dfrac{n-3}{\eta} \right) ^\beta \right] <0$ as its argument is positive. 
\\ Thus:
\begin{equation*}
2-\exp\left[   \left(  -\dfrac{n-1}{\eta} \right) ^\beta + 2 \left( \dfrac{n-2}{\eta} \right) ^\beta - \left( \dfrac{n-3}{\eta} \right) ^\beta \right] - \exp\left[ \left(  \dfrac{n-1}{\eta} \right) ^\beta - 2 \left( \dfrac{n-2}{\eta} \right) ^\beta + \left( \dfrac{n-3}{\eta} \right) ^\beta \right] <0 .
\end{equation*}
\\ As for $1<\beta\leq 2$ and $n\geq 3$, $\left( \dfrac{n}{\eta} \right) ^\beta -2 \left( \dfrac{n-1}{\eta} \right) ^\beta +\left( \dfrac{n-2}{\eta} \right) ^\beta$ is decreasing, thus: 
\begin{equation*}
-\exp\left[   \left( \dfrac{n}{\eta} \right) ^\beta -2 \left( \dfrac{n-1}{\eta} \right) ^\beta +\left( \dfrac{n-2}{\eta} \right) ^\beta \right] \leq -\exp\left[ - \left( \dfrac{n-1}{\eta} \right) ^\beta + 2 \left( \dfrac{n-2}{\eta} \right) ^\beta - \left( \dfrac{n-3}{\eta} \right) ^\beta \right].
\end{equation*}
Consequently, $A(n)<0$ for $1<\beta\leq 2$ and, trivially, $\lambda''(n)$ too.

\paragraph{Proof of Proposition~\ref{prop2}: L$^\infty$ convergence of Weibull-1 to continuous Weibull.}
Since 
$$
\sup_{t \in \mathbb R^+} |p(t|\beta,\eta) - f_W(t|\beta,\eta)| \leq \sup_{t \in \mathbb R^+} p(t|\beta,\eta) + \sup_{t \in \mathbb R^+} f_W(t|\beta,\eta),
$$ 
it is only needed to show that
\begin{equation}\label{eq:limits}
\lim_{\eta \to \infty} \sup_{t \in \mathbb R^+} p(t|\beta,\eta) =
\lim_{\eta \to \infty} \sup_{t \in \mathbb R^+}  f_W(t|\beta,\eta) = 0.
\end{equation}
The second equality comes from the fact that $\eta$ is a scale parameter, and that the Weibull density is bounded for $\beta \geq 1$. Hence:
\begin{eqnarray*}
\sup_{t \in \mathbb R^+}  f_W(t|\beta,\eta) 
&=&
\sup_{t \in \mathbb R^+} \frac{1}{\eta} f_W \left( \frac{t}{\eta}|\beta,1\right) \\
&=&
\frac{1}{\eta} \sup_{t \in \mathbb R^+} f_W \left( t|\beta,1\right) 
\stackrel{\eta\to\infty}{\longrightarrow} 0.
\end{eqnarray*}
The first equality in (\ref{eq:limits}) comes from the fact that, for all $t \in [n-1,n]$:
\begin{eqnarray*}
p(t|\beta,\eta) 
&=& 
\int_{n-1}^n f_W(t|\beta,\eta) dt \\
&\leq&
\sup_{t \in [n-1,n]} f_W(t|\beta,\eta),
\end{eqnarray*}
hence
$$
\sup_{t \in \mathbb R^+} p(t|\beta,\eta) \leq \sup_{t \in \mathbb R^+} f_W(t|\beta,\eta) \square
$$

\end{document}